\documentclass[aps,prl,preprint,superscriptaddress,showpacs]{revtex4}
\usepackage{graphicx}

\input epsf.sty
\begin{document}
\title{Dynamical spectral weight in
YBa$_2$Cu$_3$O$_y$ probed by x-ray absorption spectroscopy}
\author{J.-Y. Lin}
\email{ago@cc.nctu.edu.tw}
\affiliation{Institute of Physics, National Chiao Tung University,
Hsinchu 300, Taiwan}
\author{P. R. Lee}
\affiliation{Institute of Physics, National Chiao Tung University,
Hsinchu 300, Taiwan}
\author{Y. T. Liu}
\affiliation{Institute of Physics, National Chiao Tung University,
Hsinchu 300, Taiwan}
\author{Chung-Yu Mou}
\email{mou@phys.nthu.edu.tw}
\affiliation{Department of Physics, National Tsing Hua University,
Hsinchu 30043, Taiwan} \affiliation{Physics Division, National
Center for Theoretical Sciences, P.O.Box 2-131, Hsinchu, Taiwan}
\affiliation{Institute of Physics, Academia Sinica, Nankang, Taiwan}
\author{Y.-J. Chen}
\affiliation{Department of Electrophysics, National Chiao Tung
University, Hsinchu 300, Taiwan}
\author{K. H. Wu}
\affiliation{Department of Electrophysics, National Chiao Tung
University, Hsinchu 300, Taiwan}
\author{C. W. Luo}
\affiliation{Department of Electrophysics, National Chiao Tung
University, Hsinchu 300, Taiwan}
\author{J. Y. Juang}
\affiliation{Department of Electrophysics, National Chiao Tung
University, Hsinchu 300, Taiwan}
\author{T. M. Uen}
\affiliation{Department of Electrophysics, National Chiao Tung
University, Hsinchu 300, Taiwan}
\author{J. M. Lee}
\affiliation{National Synchrotron Radiation Research Center (NSRRC),
Hsinchu, Taiwan}
\author{J. M. Chen}
\affiliation{National Synchrotron Radiation Research Center (NSRRC),
Hsinchu, Taiwan}
\date{\today}
\begin{abstract}
The comprehensive study of the temperature dependent x-ray absorption spectroscopy (XAS) reveals a dynamical spectral weight $\alpha$ in YBa$_2$Cu$_3$O$_y$ (YBCO). Large spectral weight changes for both the Upper Hubbard band and the Zhang-Rice band due to dynamics of holes are experimentally found in the underdoped regime. A large value of $\alpha \geq 0.3$
is indispensable to describing XAS of YBCO with the conservation of states.
The value of $\alpha$ is linearly proportional to the pseudogap temperature in the underdoped
regime, but becomes smaller as the doping level goes to the undoped limit. Our results clearly indicate that the pseudogap is related to the double occupancy and originates from bands in higher energies.
\end{abstract}
\pacs{74.72.-h, 78.70.Dm, 71.27.+a, 74.72.Kf} \maketitle

The high-temperature superconductor(HTSC) is still a challenging subject in condensed
matter physics. The underlying mechanism for high-$T_c$
superconductivity is considered elusive by many researchers in this
field. In addition, above $T_c$ in the normal state, there exists
another energy scale due to which many unconventional phenomena
occur \cite{Sawatzky}. For example, the resistivity $\rho$ of HTSC
shows a linear temperature dependence above some characteristic
temperature $T^*$ and decreases below $T^*$ \cite{pseudogap,Ando}, indicating
that the scattering for the transport carriers experience becomes
less significant below $T^*$. The nature of PG have
been under debate for a long time and is believed by many to be the key to
the origin of high-temperature superconductivity. Another crucial factor in HTSC is the electron correlation. In contrast to be metals as predicted by band structure calculations, the undoped CuO$_2$ plane in HTSC is a Mott insulator with a strong correlation gap, $U$, separating the upper Hubbard band and the lower Hubbard band. As one dopes HTSC with holes into the oxygen orbitals, the holes hybridize with Cu spins to form the Zhang-Rice
singlet \cite{ZhangRice} band residing between the upper Hubbard band and the lower Hubbard band. It is therefore widely believed that the Hubbard model and thus the Mott physics, which dealt with the strong electron correlation, are indispensable to the understanding of the HTSC underlying mechanism. Even though the on-site Couloumb repulsion suppresses electrons to
doubly occupy the same Cu site, the double occupancy may occur mediated by the non-zero $t$/$U$ term ($t$ being the the hopping constant) and could hop off to a neighboring
empty site. Hence there must be dynamical contributions to the spectral weight. In this Letter, for the first time through the dynamical effects revealed by the temperature dependent x-ray absorption spectroscopy (XAS), a strong experimental link between PG and the double occupancy in the strong electron correlation regime is found.

One of the key features that makes doped HTSC different from
conventional metals is that the Zhang-Rice band that supports
superconductivity is not rigid and its spectral weight depends
strongly on the doping level and the upper Hubbard band. Indeed, as shown in early x-ray
absorption spectroscopy (XAS) \cite{Chen}, there are large spectral
weight changes for the Zhang-Rice band as the doping level is tuned. This spectral weight
change is attributed to the spectral weight transfer from the upper Hubbard band. In the atomic limit, for each hole, there are two ways of adding an electron into the hole. Consequently, it is expected that the weight among the the Zhang-Rice band, the lower Hubbard band and the upper Hubbard band is 2$p$, $1-p$, and $1-p$ respectively, with $p$ being the doping level. Experimentally, however, the Zhang-Rice band always appears to be much enhanced with weight more than 2$p$. Recently, it is further suggested that the Zhang-Rice singlet may even breakdown at $p\sim0.21$ in the overdoped region, probably correlated to the vanishing spectral weight in the upper Hubbard band \cite{Peets}. As for the dynamical contribution to the spectral weight, theoretical attempt is summarized in the derived sum rules \cite{weight, Zhang}. The ratio for the spectral
weight is shown to be 2$p+\alpha$, $1-p$, and $1-p-\alpha$ for the
the Zhang-Rice band, the lower Hubbard band and the upper Hubbard band respectively, where $\alpha$ is the dynamical spectral weight. In the lowest order, it is shown that $\alpha =2 K /U$ with $K$ being the average kinetic energy of the ground state with double
occupancy removed \cite{Zhang}. Very recently, a different weight ratio
with 2($p+\alpha$), $1-p-\alpha$, and $1-p-\alpha$ for the the Zhang-Rice band, the lower Hubbard band and the upper Hubbard band respectively was proposed in considering the symmetry between the occupied part of the lower and upper Hubbard bands. Here, however, $\alpha$ is identified as the density of doublon-holon bound states \cite{Phillipsreview}. Although different conclusions are
drawn, these results imply that it is possible for the weight of the upper Hubbard band
to vanish much early before the doping level hits one and may explain the observation of Ref.\cite{Peets}. To date, however, there has been no experimental effort to identify the dynamical contribution.

To serve this purpose, the temperature dependence of the O-$K$ edge spectral weight in YBa$_2$Cu$_3$O$_y$ was examined. O-$K$ XAS measurements, with a photon energy resolution of ~0.15 eV at 530 eV, were conducted at the high-energy spherical grating monochromator beamline of National Synchrotron Radiation Research Center in Taiwan. The spectra were taken at various temperatures
from $T$=300 K to 15 K. The self-absorption corrections were applied
to all spectra to correct the saturation effects. Details of the XAS
measurements were described elsewhere \cite{Chang}.
YBa$_2$Cu$_3$O$_y$ (YBCO) and nominal
Y$_{0.7}$Ca$_{0.3}$Ba$_2$Cu$_3$O$_{y}$ (YCBCO)
 thin films were prepared on (100) SrTiO$_3$ substrates using pulsed laser
deposition. The thickness of all the films was controlled to be
250 nm. The crystallinity of the films was analyzed by the X-ray
diffraction (XRD) pattern. The results revealed that all the films
were well-oriented (001) ones. The oxygen contents were controlled
by post annealing the as-prepared thin films at various temperatures
and oxygen pressures. This method has been proven to be capable of
controlling the oxygen contents of the YBCO films precisely and
reversibly\cite{Luo}. Resistivity $\rho (T)$ of the films was
measured by the standard four-probe method. To check the quality of
our samples, we show the data of $\rho$($T$) in Fig. 1. The oxygen
contents of YBCO samples are determined from the values of
$T_c$ \cite{Carrington}. The doping level is
obtained by using the formula $1-T_c/T_{cmax} = 82.6 (p-0.16)^2 $
with $T_{cmax}=92 K$ for YBCO and $T_{cmax}=84 K$ for YCBCO, respectively \cite{Tcformula}.
It is seen that these samples are comparable to those of
the high quality thin films and single crystals in the literature.
\begin{figure}[h]
{\includegraphics*[width=9.0cm]{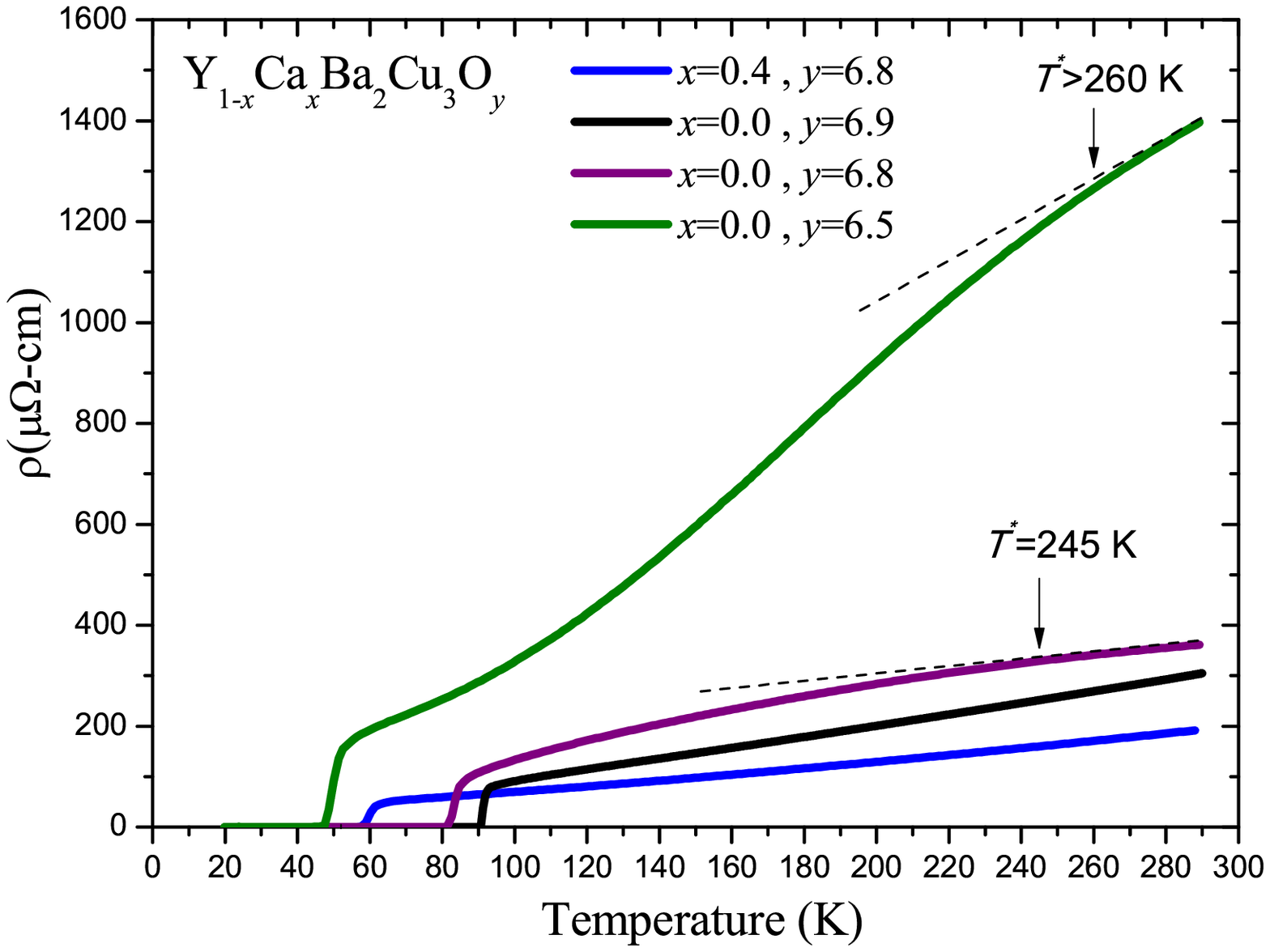}} \caption{Resistivity
$\rho$($T$) of (a) YBa2Cu3O$_{6.5}$; (b) YBa2Cu3O$_{6.8}$; (c)
YBa2Cu3O$_{6.95}$; (d) Y$_{0.7}$Ca$_{0.3}$Ba$_2$Cu$_3$O$_{6.8}$.
Here $T^*$ is the onset temperature of pseudogap.}
\end{figure}
Furthermore, the pseudogap temperature $T^*$ can be easily
identified. For the $p=0.165$ sample near the optimal doping, $T^*$
is in the vicinity of the superconducting fluctuation regime and is
difficult to determine by $\rho (T)$. However, it is plausible to
assume that $T^* \approx 100$ K at this doping.

O $K$-edge XAS of cuprates has been studied before
\cite{Chen,Merz2}. However, to the best of our knowledge, only a limited
study of the temperature effects on XAS of cuprates was reported
\cite{Hirai}. To explore $\alpha$ through the spectral weight
changes with $T$, O $K$-edge XAS of YBCO and YCBCO was measured from
$T$=300 K down to $T\leq$20 K.  Fig. 2(a)-(e) shows spectra of YBCO
and YCBCO for various doping levels. The features of the O $K$-edge XAS
at $E=527.5$ eV, $528.2$ eV, and $529.4$ eV are attributed to the
CuO chain holes, the Zhang-Rice band, and the upper Hubbard band, respectively. For YCBCO with
$p$=0.23, the feature of the upper Hubbard band is elusive as reported in
Ref\cite{Peets}.  In Fig. 2(f), the O $K$-edge XAS of SrTiO$_3$ is
shown for comparison. The spectra of SrTiO$_3$ are shown to demonstrate
the presumably trivial temperature dependence as the peaks become
sharper at low $T$. It is then clear that the temperature dependence
of O $K$-edge in YBCO is highly nontrivial. Both the
the Zhang-Rice band and the upper Hubbard band features bear noticeable changes due to the temperature
variations from $T$=300 K to $T\lesssim$20 K. Generally, the the Zhang-Rice band
spectral weight $S_{ZR}$ increases at low $T$ while that of the upper Hubbard band
($S_{UHB}$) decreases. These changes are especially obvious for the
deeply underdoped sample YBa$_2$Cu$_3$O$_6.5$. To the contrary,
temperature variations of YCBCO O $K$-edge XAS are very similar to
those of STO, mainly showing trivial line width sharpness with
decreasing $T$. To analyze the spectrum changes more quantitatively,
XAS intensity differences $\Delta S (T) \equiv S(T)-S(300 K)$
were plotted in Fig. 3. These plots further confirm the above
mentioned spectrum changes of both the the Zhang-Rice band and the upper Hubbard band features. The Gaussian peak fit was used to extract the values of $S_{UHB}$ and
$S_{ZR}$ form the $\sigma$($T$=300 K) for each YBCO. The values of
$\Delta S_{UHB}$ and $\Delta S_{ZR}$ were obtained from the
integration of the spectral weight changes as depicted in the bottom of
Fig. 2(a).
\begin{figure}[h]
{\includegraphics*[width=9.0cm]{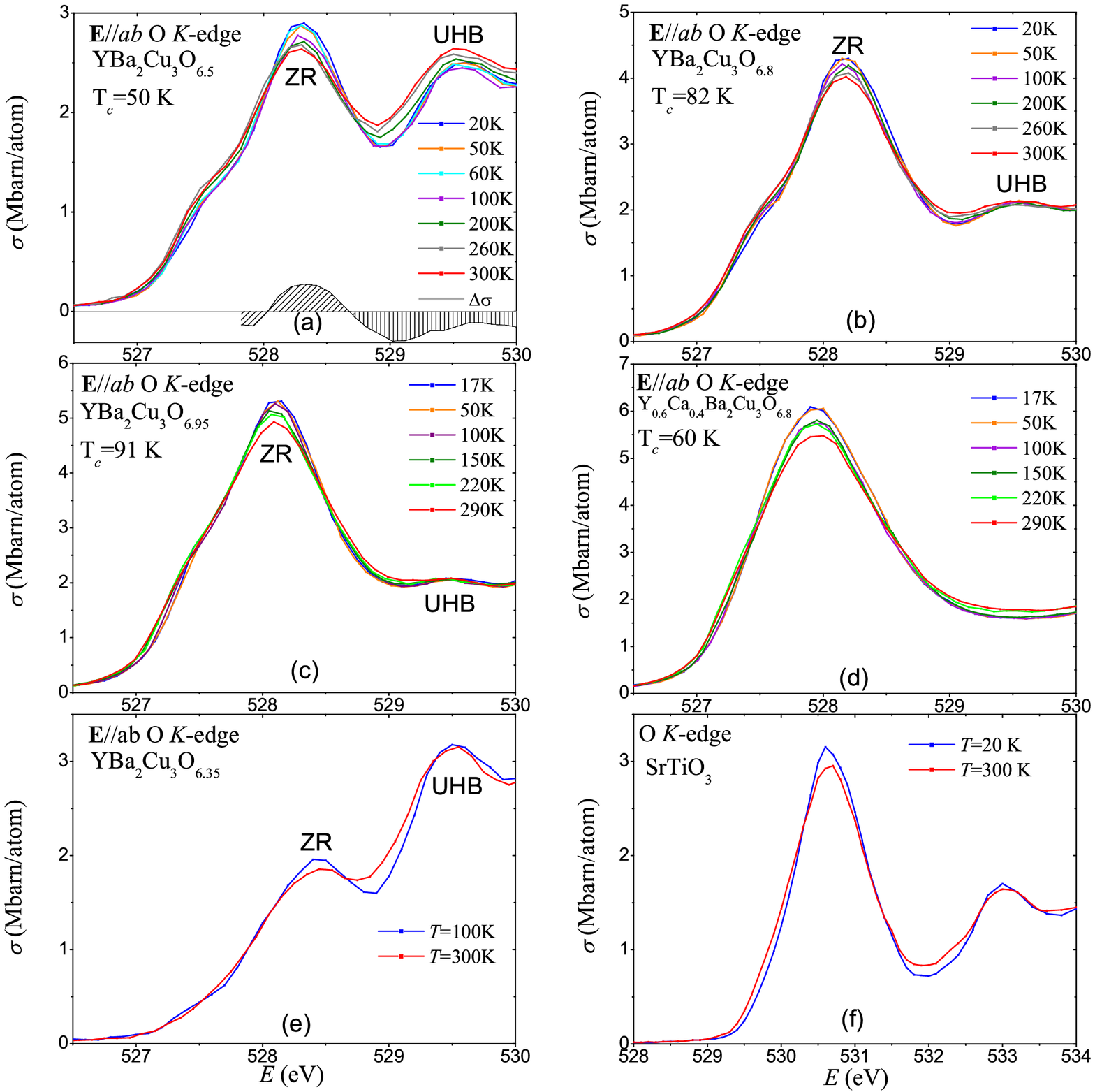}} \caption{O $K$-edge XAS
at various temperatures of (a) YBa$_2$Cu$_3$O$_{6.9}$; (b)
YBa$_2$Cu$_3$O$_{6.8}$; (c) YBa$_2$Cu$_3$O$_{6.5}$; (d)
Y$_{0.7}$Ca$_{0.3}$Ba$_2$Cu$_3$O$_{6.8}$; (e) SrTiO$_3$
(f)YBa$_2$Cu$_3$O$_{6.35}$. In (a), XAS intensity deference of
YBa2Cu3O$_{6.5}$ is denoted as the black lines. $\Delta \sigma
=\sigma (\lesssim 20 K)-\sigma (300 K)$. The areas with the slashes
and the vertical lines denote the spectral weight changes of the Zhang-Rice band and
the upper Hubbard band, respectively.}
\end{figure}
\begin{figure}[h]
{\includegraphics*[width=9.0cm]{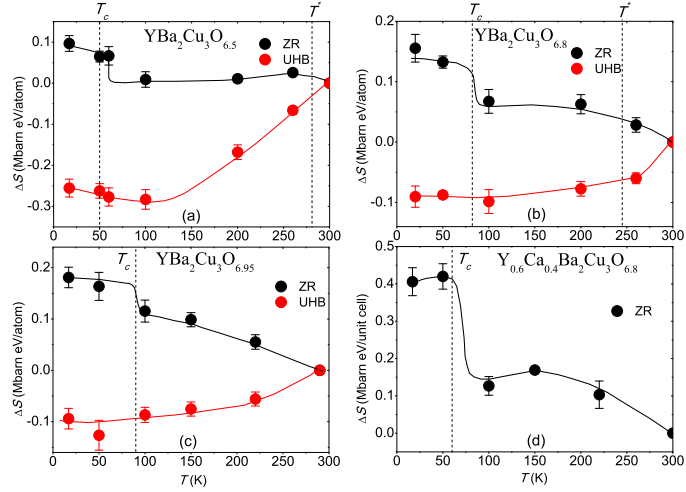}} \caption{$\Delta
S_{UHB}(T)$ and $\Delta S_{ZR}(T)$ of (a) YBa$_2$Cu3O$_{6.5}$; (b)
YBa2Cu3O$_{6.8}$; (c) YBa$_2$Cu3O$_{6.95}$; (d)
Y$_{0.7}$Ca$_{0.3}$Ba$_2$Cu$_3$O$_{6.8}$. $\Delta$$S$(300
K)$\equiv$0. $\Delta S_{UHB}(T)$ of
Y$_{0.7}$Ca$_{0.3}$Ba$_2$Cu$_3$O$_{6.8}$ due to the absence of UHB.
The lines are guides for eyes.}
\end{figure}

The non-trivial changes in the spectral weight can be related to
$\alpha$ in at least two scenarios. One scenario is to attribute the
entire $\Delta S$ to $\alpha$. In this case, $\Delta S_{UHB}/
S_{UHB}= \alpha_{UHB}/(1-p)$. This way the dynamical spectral
factor $\alpha_{UHB}$ can be estimated from the change of the upper Hubbard band.
Values of $\Delta S_{UHB}/S_{UHB}$ can be found in Table I. The
resultant values of $\alpha_{UHB}$=0.043, 0.043, and 0.10 for
$p$=6.95, 6.8, and 6.5, respectively. These numbers coincide with
those of the recent calculations with $\alpha_{UHB}$=0.05 or 0.06
for the underdoped cases \cite{Phillipsreview}. It is noted that the
formation of the proposed 2$e$ boson bound state \cite{Phillipsreview} at low temperatures
could qualitatively explain both the temperature dependence of O
$K$-edge XAS of YBCO and why the change of the upper Hubbard band is most significant
in the deeply underdoped sample (Fig. 2). However, applying the same
analysis to the changes of the Zhang-Rice band by $\Delta S_{ZR} / S_{ZR} =
\alpha_{ZR}/p$ leads to values of $\alpha_{ZR}$ which are too small:
$\alpha$ =0.008, 0.006, and 0.003 for $p=6.95, 6.8,$ and $6.5$,
respectively. The conventional picture of 2$p$+$\alpha$ would only
double the values of $\alpha_{ZR}$. Apparently, the values of
$\alpha_{UHB}$ are not consistent with those of $\alpha_{ZR}$ in
this scenario. It has been argued that the spectral weight from the associated the Zhang-Rice peak energy range might not be a good indication of $S_{ZR}$ \cite{PhillipsZRB} due to the observed saturation in the overdoped regime \cite{Peets}. In this case, the dynamical corrections are more faithfully manifested by $\alpha_{UHB}$. Since undoped cuprates are actually charge transfer insulators, it is likely that the carrier doping in O 2$p$ states and the dynamical double occupancy in Cu 3$d$ states respectively lead to different spectral weight transfers.

As another scenario considering both the conservation of states and the temperature
dependent spectral weight, $\alpha$ could already exist
at $T$=300 K and be temperature dependent. This scenario leads to the
coupled equations $\Delta S_{UHB} / S_{UHB}= \Delta
\alpha/(1-p-\alpha-\Delta \alpha)$ and $\Delta S_{ZR} /S_{ZR}=\Delta
\alpha/(2p+\alpha+\Delta\alpha)$. Apparently, these equations are
free from the dependence on electron-photon matrix elements. The
resulting $\alpha^*$ at $T^*$ and $\Delta\alpha= \alpha
(300K)-\alpha(\lesssim 20K)$ are listed in Table I. The temperature
variation of $\Delta \alpha$ is about 0.03 and is of the same order
of magnitude as $\alpha_{UHB}$ in the previous paragraph. Values of
$\alpha$ for all samples are quite large. (The 2$p$+2$\alpha$
scenario \cite{Phillipsreview} would make $\alpha$ slightly larger but do $\Delta\alpha$ about the same.) However, these large values are not implausible. In
the effective one band Hubbard model for cuprates, $U$ is actually
the charge transfer gap between O 2$p$ and Cu 3$d$ states. From the
energy difference between the Zhang-Rice and the upper Hubbard band peaks in Fig. 2, $U\approx$1.5
eV, which is roughly the same as that ($\leq$2.0 eV) obtained by
optical measurements \cite{Leereview}. This small $U$ is responsible
for the observed large exchange energy $J$=0.13 eV \cite{Leereview} and also leads to a large $\alpha$. Since the measured $K$ for
optimally doped cuprates \cite{Molegraff} is 0.345 eV/Cu, using
$\alpha =2K/U$, the estimated $\alpha \sim 0.46$ is consistent with
those obtained in Table I. Furthermore, the increase in $K$ across
$T_c$ due to coherence of electrons also leads to estimates of
$\Delta \alpha$ comparable to those in Table I. The large $\alpha$
could account for the faster growth of $S_{ZR}$ than 2$p$ (or the
faster decay of $S_{UHB}$ than 1-$p$), as suggested in Ref.
\cite{weight}. It could also, at least partially, be responsible
for the observed vanishing of the the upper Hubbard band peak in the deeply overdoped
cuprates (Fig. 2(d)) in addition to the factor of the Zhang-Ricesinglet
breakdown \cite{Peets}. Data in Table I clearly show that $\alpha$
is linearly proportional to $T^*$ in the underdoped regime and
eventually bends over and decreases in the undoped limit
as shown by sample with $y=6.35$. It indicates that the formation of
the pseudogap enhances the particle addition states in cuprates \cite{Phillipsreview}. Here the mechanism is quite similar to that for the decrease in resistivity
when $T$ goes below $T^*$.


\begin{table}
\caption{Parameters extracted from Fig. 3. Here for samples with
superconductivity in low temperature ($y=6.9,6.8,6.5$), $\Delta
\alpha $ is the difference of $\alpha$ at $T=300K$ and $T \lesssim
20K$. $\alpha^*$ is the value of $\alpha$ at $T^*$. We find $\alpha
\approx 0.0011T^*$. Note that for $y=6.35$, there is no measurable
$T^*$ and ($\alpha^*$, $\Delta \alpha$) are estimated based on
$T=300K$ and $T=200K$. \label{data}}
\begin{ruledtabular}
  \begin{tabular}{ccccccc}
    Sample & $p$  & $T^*$ & $\Delta S_{RZ}/S_{RZ}$ & $\Delta S_{UHB}/S_{UHB}$ & $\alpha^*$ & $\Delta \alpha$ \\ \hline
    YBa$_2$Cu$_3$O$_{6.9}$& 0.163 & 290 & 0.04724 & 0.05218 & 0.435 & 0.029 \\
    YBa$_2$Cu$_3$O$_{6.8}$ & 0.13 & 240 & 0.04695 & 0.04932 & 0.587 & 0.027 \\
    YBa$_2$Cu$_3$O$_{6.5}$ & 0.085 & 100 & 0.03661 & 0.10985 & 0.644 & 0.03 \\
    YBa$_2$Cu$_3$O$_{6.35}$ & 0.06 & * & 0.02138 & 0.02968 & 0.496 & 0.013 \\
  \end{tabular}
\end{ruledtabular}
\end{table}

One of the interesting questions is whether the temperature
variation of the spectral weight correlates with the formation of the pseudogap. This issue was previously raised by an attempt to integrate the high energy scale in the Hubbard model \cite{Phillipsreview}. In Fig. 3(a) and (b), the pseudogap temperatur $T^*$ is indicated for
$p$=6.5 and 6.8. $T^*$ of the slightly overdoped $p$=6.95 sample is
difficult to define from $\rho (T)$. In Fig 3(b), the rapid decrease
in $S_{UHB}$ with decreasing $T$ occurs around $T^*$ suggesting a
correlation between $\Delta$$S_{UHB}$ and $T^*$. However, this
correlation is not so obvious either in Fig. 3(a) or 3(c). Likely,
the decrease in $S_{UHB}$ is a crossover rather than a well defined
transition. On the other hand, a jump of $S_{ZR}$ around $T_c$ with
decreasing $T$ was observed for all samples with superconducting
transitions. This jump could be a manifestation of the charge
redistribution in YBCO because it is energetically favorable to have
more holes residing in the CuO$_2$ planes below $T_c$. This was
actually predicted in Ref. \cite{Khomskii} with the change being
estimated to be of order $\Delta/ E_F$, where $\Delta$ is the
superconducting gap and $E_F$ the Fermi energy. This change of
carrier concentration can be up to several percents in YBCO
\cite{Khomskii} and is comparable to the present magnitudes. These
jumps appear to occur slightly above $T_C$, indicating that
condensation energy sets in before the coherence of the cooper
pairs. The significant larger change in the overdoped sample
compared with other three samples clearly reflects the electronic
structure change from underdoping to overdoping.

We note in passing that a similar $T$-dependent O $K$-edge XAS was
found in La$_{1-x}$Sr$_{1+x}$MnO$_4$. Here the mechanism was ascribed to the
tetragonal distortion \cite{Merz}, characterized by the bond length
ratio, $D$=Mn-O(apical)/Mn-O(in-plane). Due to that $D$ increases
with decreasing $T$, the hybridization of O 2$p_{x,y}$ (2$p_{z}$)
states with Cu 3$d$ states leads to the observed $T$-dependence.
However, in contrast to the case of La$_{1-x}$Sr$_{1+x}$MnO$_4$, $D$ deceases
with $T$ in YBCO\cite{Sharma}. Consequently, the observed $T$
dependence in the present experiments is not due to the charge
redistribution induced by the lattice distortion.

In summary, in addition to using carrier doping, it is verified that the temperature variation leads to further spectral weight transfer of O $K$-edge XAS in YBCO.
The key experimental features of the present work bear the non-trivial $T$ dependence of XAS in YBCO. These spectral changes associated with the upper Hubbard band and the Zhang-Rice band are identified as the dynamical contribution. The observation that $\alpha$ is linearly proportional to $T^*$ in the underoped regime indicates a strong link between pseudogap and the high energy states through the double occupancy. Scenarios like 2$e$ bosons could qualitatively explain both the temperature dependence of the spectra and why the changes are most significant in the deeply underdoped sample \cite{Phillipsreview}. However, The details of these changes could not be accounted by any existing theoretical model.

We thank T. K. Lee, and P. Phillips for useful discussions. This work
was supported by the National Science Council of Taiwan.

\end{document}